\documentclass{article}

\usepackage{pifont}

\usepackage[linesnumbered,lined,ruled,commentsnumbered]{algorithm2e}

\usepackage{spconf,amsmath,graphicx}
\usepackage{bm}
\usepackage{multicol}
\usepackage{multirow}

\usepackage{enumitem}
\usepackage{url,hyperref,cleveref}
\usepackage{diagbox} % 创建具有斜线表头的表格单元格 \diagbox{Col 1}{Col 2}.
\usepackage{booktabs} % \toprule \midrule \bottomrule \cmidrule{start-end} 表格的 start 到 end 列添加粗线

\usepackage{subfigure}

\title{Title}
\title{TDT-KWS: Fast and accurate keyword spotting using Token-and-Duration Transducer}
\name{Yu Xi$^1$, Hao Li$^2$, Baochen Yang$^1$, Haoyu Li$^1$, Hainan Xu$^3$, $^{\dagger}$Kai Yu$^1$
\thanks{$^{\dagger}$Kai Yu is the corresponding author.}}
\address{$^1$MoE Key Lab of Artificial Intelligence, AI Institute, X-LANCE Lab, Shanghai Jiao Tong University
\\$^2$AISpeech Ltd, Suzhou, China
\\$^3$NVIDIA, U.S.A.}

\begin{document}
\ninept
\maketitle

%\include{content/0_abstract}

% Body
% abstract
\linespread{0.8}
\begin{abstract}
% Designing an efficient keyword spotting (KWS) system that delivers exceptional performance on resource-constrained edge devices has long been a subject of significant attention. Existing KWS search algorithms typically follow a {\em frame-synchronous} approach, where search decisions are made repeatedly at each frame despite the fact that most frames are keyword-irrelevant. In this paper, we propose TDT-KWS, which leverages token-and-duration Transducers (TDT) for KWS tasks. We also propose a novel KWS task-specific decoding algorithm for Transducer-based models, which supports highly effective and efficient {\em frame-asynchronous} keyword search in streaming speech scenarios. With evaluations conducted on both the public Hey Snips and self-constructed LibriKWS-20 datasets, our proposed KWS-decoding algorithm produces more accurate results than conventional ASR decoding algorithms, while running over 200X faster; Additionally, TDT-KWS achieves on-par or better wake word detection performance than both RNN-T and traditional TDT-ASR systems. Furthermore, experiments show that TDT-KWS is more robust to noisy environments compared to RNN-T KWS. 

Designing an efficient keyword spotting (KWS) system that delivers exceptional performance on resource-constrained edge devices has long been a subject of significant attention. Existing KWS search algorithms typically follow a {\em frame-synchronous} approach, where search decisions are made repeatedly at each frame despite the fact that most frames are keyword-irrelevant. In this paper, we propose TDT-KWS, which leverages token-and-duration Transducers (TDT) for KWS tasks. We also propose a novel KWS task-specific decoding algorithm for Transducer-based models, which supports highly effective {\em frame-asynchronous} keyword search in streaming speech scenarios. With evaluations conducted on both the public Hey Snips and self-constructed LibriKWS-20 datasets, our proposed KWS-decoding algorithm produces more accurate results than conventional ASR decoding algorithms. Additionally, TDT-KWS achieves on-par or better wake word detection performance than both RNN-T and traditional TDT-ASR systems while achieving significant inference speed-up. Furthermore, experiments show that TDT-KWS is more robust to noisy environments compared to RNN-T KWS. 
\end{abstract}


\section{Introduction}
\label{sec:intro}
\emph{Keyword spotting} (KWS) is the task of detecting predefined keywords within streaming audio~\cite{guoguo-dnnkws}. Due to the rapid development of the Internet of Things (IoT) and intelligent cockpits, KWS systems, in particular \emph{wake word detection} (WWD) systems, are now widely used in various aspects of our daily lives~\cite{KWS_IOT}. In order to provide a seamless human-machine interaction experience, it is essential to find the right balance between minimizing false alarms, maximizing recall (the ability to detect the keywords correctly), and ensuring low computational burden for small-footprint KWS.
%~\cite{kws_recall}

In recent years, RNN-T \cite{RNNT}, also known as Transducers, has achieved great success in automatic speech recognition (ASR) ~\cite{RNNT-ASR_01, RNNT-ASR_02, RNNT-ASR_03, RNNT-ASR_04, RNNT-ASR_05}, speech translation (ST)~\cite{RNNT-ST_01}, and KWS~\cite{RNNT-KWS_01, RNNT-KWS_02, RNNT-KWS_03-RNNTByteDance, RNNT-KWS_04, RNNT-KWS_05_CaTT-KWS}. Since WWD models are typically deployed on edge devices, it is necessary to design a lightweight neural network to account for the limited computational and storage resources. \emph{Tiny Transducer}~\cite{Tiny-RNNT}, a phoneme-based Transducer, is proposed to address the problem of on-device streaming speech recognition. It consists of deep feedforward sequential memory network (DFSMN)~\cite{DFSMN} blocks as the Transducer encoder, a stateless network as the predictor, and a linear layer as the joiner to reduce the network's parameters. 
Subsequent work on small-footprint WWD~\cite{RNNT-KWS_03-RNNTByteDance} and CaTT-KWS~\cite{RNNT-KWS_05_CaTT-KWS} both follow the same network architecture as Tiny Transducer. In this paper, we follow part of the Tiny Transducer's configuration as well.  

Despite their superior performance, the auto-regressive decoding of Transducers is computationally intensive and can introduce significant computational latency, especially for tasks like KWS running under limited hardware resources. Recently, \emph{Token-and-Duration Transducer} (TDT) ~\cite{TDT} is proposed to alleviate this issue by jointly predicting a token and its duration. TDT achieves better performance and significant inference speed-up compared to the original RNN-Ts in a number of sequence modeling tasks. The refined design and advantages of the TDT model make it suitable for Transducer-related KWS systems, which is the focus of this paper.

Note, that there is a straightforward method to perform KWS with a Transducer model, by simply running conventional ASR decoding on the audio sequence, and then checking if the keyword is in the decoding output. To our best knowledge, all existing Transducer-based KWS systems~\cite{RNNT-KWS_01, RNNT-KWS_02, RNNT-KWS_03-RNNTByteDance, RNNT-KWS_04, RNNT-KWS_05_CaTT-KWS} follow this approach. However, this use of Transducers for KWS is not optimal or efficient, since in conventional ASR decoding, the search space is not constrained to the specified keyword, and the search algorithm is not tailored to KWS tasks. In this paper, we propose a novel \emph{KWS decoding algorithm} specifically designed for KWS tasks, which constrains the search space for fixed keywords and can fully realize the potential of TDT for KWS. This paper makes the following contributions:

% We point out that, it is a non-trivial task to efficiently and effectively apply TDT to KWS. Most KWS systems, including Transducer-based KWS systems, employ a two-stage architecture including an acoustic model and a search or decoding module. To the best of our knowledge, all existing Transducer-based KWS approaches\cite{all_rnnt_kws} rely on conventional \emph{ASR decoding algorithms} to detect keywords, no matter whether they are used for fixed or customizable KWS. However, in conventional ASR decoding algorithms, the search space is not constrained to the specific keyword, which may lead to sub-optimal performance for KWS tasks and is not ideal. In this paper, we propose a novel \emph{KWS decoding algorithm} specifically designed for KWS tasks, which constrains the search space for fixed keywords and can fully realize the potential of TDT for KWS. This paper makes the following contributions:

\setlist[itemize]{leftmargin=0.5cm}
\begin{itemize}
    \item We propose a novel KWS decoding algorithm that dynamically detects the start of keywords for Transducers in streaming continuous speech. The proposed algorithm can obtain better KWS performance compared to conventional ASR decoding algorithms.
    %, also bringing a relative speed-up of over 200X.
    \item We propose TDT-KWS, which achieves comparable or better performance on the open-source KWS dataset ``Hey Snips"~\cite{Snips} and our self-constructed LibriKWS-20 dataset derived from LibriSpeech~\cite{LibriSpeech} compared to conventional RNN-T KWS, while running an additional 2-4 times faster during inference.
    \item TDT-KWS demonstrates greater robustness than conventional RNN-T systems in low signal-to-noise ratio (SNR) environments, which is crucial for KWS to reduce false alarms in extreme environments.
\end{itemize}

\def\smallPlus{\texttt{+}}
\def\smallMinus{\texttt{-}}
\renewcommand{\algorithmcfname}{Alg.}

\section{TDT Based Keyword Spotting}
\label{sec:method}

% In this section, we firstly present the overview of Transducer KWS systems. Subsequently, we introduce the principle and the architecture of TDT briefly. The proposed streaming decoding algorithm for RNN-T KWS system and the proposed streaming decoding algorithm for TDT are introduced in the end.

\begin{figure}[t]
  \centering
    \includegraphics[width=\linewidth]{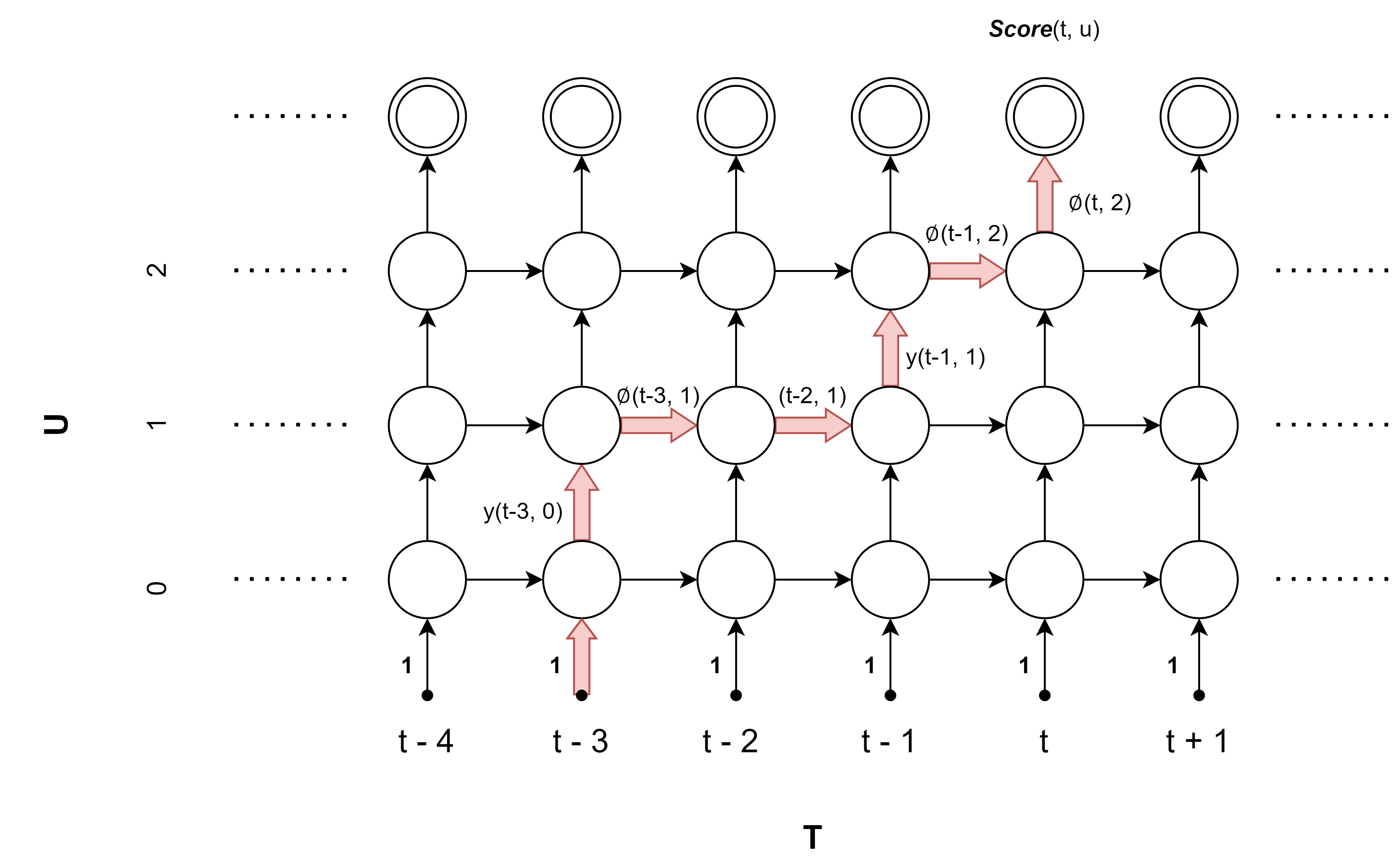}
    \vspace{-1.5em}
    \caption{Decoding path for RNN-T KWS System. Each node $(t, u)$ represents the highest score obtained by outputting the first $u$ elements of the keyword up to time $t$. The horizontal arrow originating from node $(t, u)$ indicates the probability $\phi(t, u)$ of outputting {\it blank}. The vertical arrow represents the probability $y(t, u)$ of outputting the $(u+1)$-th element of the keyword at time $t$. To identify the optimal path for the keyword at time $t$, the path with the maximum score is illustrated by red arrows. This path corresponds to the most probable sequence of the keyword at time t.}
    \label{fig:rnnt-kws-decoding-lattice}
    \vspace{-1.5em}
\end{figure}

\subsection{Transducers}
\label{sec:2.1}
% Transducers consist of an encoder, a predictor, and a joiner (also known as a joint network). The encoder takes acoustic features as input, representing the acoustic properties of the speech signal. The decoder takes text input, usually in the form of linguistic units like phonemes, graphemes, or sub-words. The higher-level representations extracted by these two modules are then combined and fed into the joint network, which generates a probability distribution over the entire vocabulary, including the blank symbol $\phi$. Considering the difference of input between Transducers and other acoustic architectures, in addition to the acoustic features, the Transducer-based system requires an additional input of the partial decoding hypothesis sequence, which provides semantic information to bias the logit prediction. In the case of the KWS task, the search sequence is predetermined (i.e., the keyword), making this part of the input fixed. This differs from ASR systems, as they cannot determine the decoding sequence before inference. 

A Transducer consists of an encoder, a predictor, and a joiner (or a joint network). The encoder takes acoustic features as input, capturing the acoustic properties of the speech signal. On the other hand, the decoder accepts text input, typically in the form of linguistic units such as phonemes, graphemes, or sub-words. The higher-level representations obtained from these two modules are then merged and fed into the joiner, which generates a probability distribution $P(v|t, u)$, where $v$ can be any token in the vocabulary or a special blank symbol $\phi$, and $t, u$ refer to indices to acoustic frame and text tokens, respectively. All modules of a Transducer model are jointly trained to maximize the probability of the correct labels given the acoustic input, where the probability of the labels sums over all possible alignments of input/output, by including blanks to the label sequence.

%In contrast to other acoustic architectures, Transducer-based systems require an additional input: the partial decoding hypothesis sequence. This sequence contributes semantic information to bias the logit prediction. In the case of the KWS task, the search sequence is pre-determined, i.e., the keyword. This differs from ASR systems, where the decoding sequence cannot be determined before the inference stage. 

% We leverage the advantages offered by Transducers to a greater extent for the KWS task, rather than simply considering Transducers as superior acoustic models. Therefore, we design search algorithms for RNN-T and TDT-KWS systems. Details are introduced in \Cref{sec:section_decoding_alogs}.

% To further enhance the advantages provided by Transducers, we specifically focus on leveraging them for KWS, rather than solely regarding Transducers as superior acoustic models. Therefore, we introduce a search algorithm tailored for RNN-T KWS and TDT-KWS systems. For a more thorough understanding of these algorithms, please refer to Section \ref{sec:section_decoding_alogs}, where we provide detailed explanations and descriptions.

\subsection{Token-and-Duration Transducers}
\label{sec:2.2}

% TDT differs from the original Transducers as it not only predicts the current token, but also the token duration of current emission. We can make a simple comparison with vanilla Transducers. For RNN-T, the output probability of the joint network can be formalized as
% $P(v|t,u)$, where $v$ could either be a label token or blank $\phi$, $t$ and $u$ are the acoustic frame index and the predicted text token, respectively. But for TDT, the joint probability can be written as $P(v,d|t,u)$, with an extra variable duration $d$ at location $(t, u)$. Further, the authors in \cite{TDT} assume that the token $v$ and duration $d$ are conditionally independent:

TDT improves upon conventional Transducers by incorporating the prediction of token duration in the joiner output. While conventional Transducers predict $P(v|t,u)$, TDT model predicts a joint distribution $P(v,d|t,u)$. The additional variable $d$ represents the predicted duration at the location $(t,u)$. In \cite{TDT}, a simple conditional independence assumption was made, i.e.,

\begin{equation}
    P(v,d|t,u) = P_{T}(v|t,u)P_{D}(d|t,u),
\end{equation}
where $P_{T}(.)$ and $P_{D}(.)$ represent the token and duration output distribution, respectively.

TDT's duration prediction $d$ guides its decoding procedure to skip input frames during inference. Specifically, the max duration $\mathcal{D}_\text{max}$ is a hyper-parameter pre-defined before training. For example, if we set $\mathcal{D}_\text{max}$ to 4, the possible predicted durations would be $\mathcal{D} = \{0,1,2,3,4\}$. It is important to note that a larger preset value of $\mathcal{D}_\text{max}$ leads to more aggressive frame-skipping by the model.

Due to the introduction of duration modeling, TDT performs {\em frame asynchronous} search and consequently achieves significant speedup compared to the conventional {\em frame synchronous} Transducer-based systems, such as RNN-T. This motivates us to introduce TDT to KWS in this paper. However, we point out that the default search algorithms for conventional Transducers and TDT are designed for ASR tasks, which do not take into account the characteristics of KWS tasks and hence are neither effective nor efficient. To address this, we propose a KWS-specific decoding algorithm for Transducer models.   

\subsection{Efficient Streaming KWS-Decoding Algorithm}
\label{sec:section_decoding_alogs}

\begin{algorithm}[t]
\caption{Streaming KWS-Decoding for Transducers}
 \label{alg:alg_rnnt_tdt_algo}
 \KwIn{keyword $\textbf{y} = \{\phi, y_1, \cdots, y_U\}$}
 \KwOut{ $Score[T]$}  	
 \BlankLine
 Init: $Score[1:T] = \{0\}$, $\phi(0, u) = 0, \delta(0, u) = 1 \text{ for } 0 \le u \le U$, $G = \{\phi\}$, $t = 1$, $d = 1$.

 \While{$t\le T$}{
    $\delta(t, 0) = 1$ \;
    \For{$u\leftarrow 1$ \KwTo $U$}{

        $\delta(t, u) = \max(\delta(t, u-1) \cdot y(t, u-1),\delta(t-d, u) \cdot \phi(t-d, u))$
    }
    % should we dot \phi(t, U) or not??? 
    $Score[t] = \delta(t, U) \cdot \phi(t, U)$ \;
    
    \uIf{\text{RNN-T KWS}}{
        $d = 1$\;
    }
    \ElseIf{\text{TDT KWS}}{
        $d = \operatorname*{argmax}_{d} P_D(d|t, G)$ \;
        $v = \operatorname*{argmax}_{v} P_T(v|t, G)$ \;
        $G \leftarrow G \cup \{v\}$ \;
        \For{$i\leftarrow 1$ \KwTo $d-1$}{
            $Score[t + i] = 0$\;
        }
    }
    $t = t + d$ \;
  }
 \Return $Score[1:T]$

\end{algorithm}

%\subsubsection{Introduction and Definition}
% In this part, we introduce an efficient decoding streaming algorithm for both RNN-T and TDT KWS. Instead of recursively feeding the Transducers' output token into the predictor like ASR, we feed decoded keyword token sequence to the predictor, as we just would like to detect whether the keyword exists in the audio. This is somewhat similar to the parallel computation of the RNN-T loss or the teacher-forcing strategy used for training of auto-regressive models. 

% efficient -> effective 
This section presents an effective KWS-specific steaming decoding algorithm, which works for both conventional Transducers and TDT, and could be extended to other extensions of Transducers as well. Unlike ASR decoding where the predictor is recursively fed \emph{partial hypotheses} during decoding, we feed only the \emph{decoded keyword token sequence} to the predictor, since the main objective of KWS is to detect the presence or absence of the keyword in the audio rather than generating a complete hypothesis. This approach is somewhat similar to the parallel computation of the RNN-T loss or the teacher-forcing strategy commonly used to train auto-regressive models. We denote acoustic features as $\mathbf{x} = \{x_0, x_1, x_2, \dots, x_T\}$ and keyword token sequence $\mathbf{y} = \{y_0=\phi, y_1, y_2, \dots, y_U\}$ ($y_0$ denotes the blank symbol), and we represent the token/blank emission probabilities following standard Transducer literature as follows:
\begin{equation}
y(t, u) = P(y_{u+1}| \mathbf{x}_{[1:t]}, \mathbf{y}_{[0:u]}),
\end{equation}
and
\begin{equation}
\phi(t, u) = P(\phi| \mathbf{x}_{[1:t]}, \mathbf{y}_{[0:u]}),
\end{equation}
for $t \in \left[0, T\right]$ and $u \in \left[0, U\right]$.

For conventional Transducer models, the decoding algorithm employs a decoding lattice as depicted in \Cref{fig:rnnt-kws-decoding-lattice}. We define $\delta(t, u)$ as the path with the highest score among all paths reaching the node $(t, u)$. In a streaming decoding scenario, where the keyword can start at any moment within the speech stream, we need to pay special attention to the starting time step of the keyword. To address this, we assign a score of 1 to $\delta(t, 0)$ for each time step $t$, which allows for detecting the keyword starting from any moment and seamlessly facilitates wake word detection in continuous speech. By leveraging dynamic programming, as shown in \Cref{alg:alg_rnnt_tdt_algo}, we can efficiently compute the complete path score $\delta(t, U)$ of the keyword. Subsequently, the keyword confidence at time $t$ can be obtained by multiplying the path score $\delta(t, U)$ and the blank score $\phi(t, U)$:

\begin{equation}
Score[t] = \delta(t, U) \cdot \phi(t, U).
\end{equation}

%\subsubsection{Some Additional Searching Details for TDT-KWS}
% TDT predicts not only the distribution of the next token but also the duration of the token at decoding status $(t, u)$. We approximate the number of frames skipped at the current decoding timestep $t$ using the predicted duration obtained from the greedy search. We choose the duration predicted by greedy search for two reasons: (1) Greedy search can provide accurate enough frame-skipping information. (2) The inference speed of greedy search is fast and imposes minimal additional computational overhead. Besides the duration prediction, other part of the decoding algorithm is identical to the original Transducers. The complete implementation can be found in \Cref{alg:alg_rnnt_tdt_algo}. 

% To estimate the number of frames skipped at the current decoding timestep $t$ for TDT-KWS, we utilize the predicted duration obtained from the greedy search. Greedy search is chosen for duration prediction due to two reasons: (1) It provides accurate enough frame-skipping information. (2)The inference speed is fast, adding minimal additional computational overhead. Apart from the duration prediction, the decoding algorithm remains identical to the original Transducers. You can find the complete implementation of the algorithm in \Cref{alg:alg_rnnt_tdt_algo}.

% For TDT models, small changes to the algorithm are needed for the KWS-decoding algorithm, since the predicted duration allows us to skip frames. Details of the change are shown in \Cref{alg:alg_rnnt_tdt_algo}.

For TDT models, small modifications are required in the KWS decoding algorithm, as the predicted duration allows us to skip frames. The specific details are presented in \Cref{alg:alg_rnnt_tdt_algo}

% Greedy search is chosen for duration prediction for two reasons: (1) It provides accurate enough frame-skipping information. (2) The inference speed is fast, adding minimal additional computational overhead. Apart from the duration prediction, the decoding algorithm remains identical to the original Transducers. 

\section{experimental setup}
\label{sec:exp}
\subsection{Datasets}

We evaluate TDT-KWS on three scenarios: 1. fixed single keyword utterance, 2. keywords in continuous speech, and 3. keywords in noisy environments. We use the following datasets.
\begin{itemize}
    \item \textbf{Hey Snips} \cite{Snips}. The Hey Snips dataset is an open-source KWS dataset that specifically uses ``Hey Snips'' as the keyword, pronounced without pause between the two words. The dataset consists of 5876, 2504, and 2588 positive utterances, and 45344, 20321, and 20821 negative utterances in the train, dev, and test datasets. Due to the absence of complete transcripts for the negative utterances, we exclusively utilize the negative segments to evaluate false alarms rather than include them in the training process. To create the false alarm dataset, we combine all the negative utterances from the train, dev, and test datasets, resulting in a dataset with approximately 97 hours of audio. 
    \item \textbf{LibriSpeech} \cite{LibriSpeech}. LibriSpeech is a widely utilized speech corpus that comprises 960 hours of read English speech accompanied by corresponding transcripts. In order to simulate the scenario of detecting wake words within a continuous stream of speech, we choose 20 specific words from the LibriSpeech dataset to serve as keywords, forming a KWS version of LibriSpeech, known as \textbf{LibriKWS-20}. To construct the false alarm dataset, we combine the audio samples from the test dataset that do not contain the selected keywords. The duration of the false alarm datasets for both the test-clean and test-other test sets is about 3 hours.
    We present these selected keywords in \Cref{table:librikws-20-keywords}.
    % LibriSpeech dataset is a widely used speech corpus which contains 960 hours read English speech with corresponding transcripts. To simulate the scenario of detecting wake words in a continuous stream of speech, we select 20 words from the dataset serving as keywords to construct the KWS version of LibriSpeech (\textbf{LibriKWS-20}). The audios from the test dataset that do not contain the selected keywords are combined to create the false alarm dataset. The duration of false alarm datasets for test-clean and test-other are both about 3 hours. 
    %We present these selected keywords in \Cref{table:librikws-20-keywords}.
    \item \textbf{WHAM!} \cite{WHAM}. The WHAM! dataset is an ambient noise corpus recorded in urban environments such as restaurants, bars, etc. It comprises various scenarios and provides a collection of background noises encountered in real-world settings. To evaluate the robustness of the model in noisy environments, we mix the test portion of the WHAM! dataset with the positive audio samples from the Hey Snips dataset at different SNRs.
\end{itemize}

\subsection{Experimental Setup}
%2,257,608 2,258,893
During training, we incorporate online speech perturbation~\cite{Speed_Perturbation}, where the warping factors are randomly selected from the set ${0.9, 1.0, 1.1}$.
The acoustic features consist of 40-dimensional log Mel-filter bank coefficients (FBank) extracted using a 25ms window with a 10ms window hop. SpecAugment ~\cite{Specaug} is applied during training, employing a maximum frequency mask range of $F=10$ and a maximum time mask range of $T=50$. Specifically, two masks of each type are used for each data sample. We splice five frames from the left and right contexts to construct 440-dimensional features and set the frame-skipping parameter to 3, resulting in three times subsampling. We follow Tiny Transducer~\cite{Tiny-RNNT} in the encoder architecture and the setup of hyper-parameters. Specifically, The encoder comprises 6 DFSMN layers, with hidden and projection sizes of 512 and 320, respectively. For all DFSMN layers, the left and right context frames are set to 8 and 2. We utilize the stateless predictor implemented in NeMo~\cite{NeMo}, configuring the context sizes as two and the embedding dimensions as 320. The joiner converts the 320-dimensional encoder and decoder outputs into 256-dim representations, which are activated and projected to final outputs. The final output units for the original Transducer include 70 monophones, derived from the CMU pronouncing dictionary ``cmudict-0.7b" \cite{cmudict}, and one blank symbol. For TDT, the output units also include the possible predicted duration options. Both RNN-T and TDT modes have approximately 2M parameters suitable for on-device KWS systems. The additional parameters in TDT for duration prediction can be negligible, as they only account for 0.1\% of the total parameter count.

\begin{table}[t]
  \centering
  \newcolumntype{S}{>{\small}c}
  \begin{resizebox}{1.0\columnwidth}{!}
  {
    \begin{tabular}{S S S S S}
      \toprule
      ~& ~&\multirow{2}*{\textbf{Keywords}}&~&~ \\
      ~& ~&~&~&~ \\
      \midrule
      almost & anything & behind & captain & children \\
      company & continued & country & everything & hardly \\
      himself & husband & moment & morning & necessary \\
      perhaps & silent & something & therefore & together \\
      \bottomrule
    \end{tabular}%
   }
   \end{resizebox}
   \linespread{0.9}
   \caption{The selected keywords in LibriKWS-20. The keywords for test-clean and test-other two datasets are the same.}
  \label{table:librikws-20-keywords}
% \vspace{-7pt}
\end{table}

For LibriKWS-20, we train models using LibriSpeech-960h. For Snips, we initialize the training with the pre-trained model on LibriSpeech-960h. Then, we train the model using the positive data from Hey Snips, along with an equally sized set of utterances randomly selected from LibriSpeech-960h. 

\begin{table}[b]
  \centering
  \newcolumntype{S}{>{\small}c}
  \begin{resizebox}{1.0\columnwidth}{!}
  {
    \begin{tabular}{ S S S S S }
      \toprule
      \multirow{2}*{\textbf{Model}} & \multirow{2}*{\textbf{Decoding Alg.}} & \multicolumn{3}{c}{\textbf{Macro-recall}} \\
      \cmidrule(lr){3-5}
      ~ & ~ & Snips & test-clean & test-other \\
      \midrule
       \multirow{3}*{RNN-T KWS}  & Greedy Search & 79.7 & 82.5 & 57.6  \\ 
         ~& Beam Search (beam=10) & 89.4 & 82.7 & 60.0 \\
        % ~& Greedy Search & 79.7 & 82.5 & 57.6 & 111X\\
        ~& Proposed & \textbf{98.7}  & \textbf{97.5} & \textbf{88.1}  \\
      \midrule
      
      \multirow{2}*{TDT-KWS} & Greedy Search & 85.4 & 84.6 & 63.2 \\
        ~& Proposed & \textbf{98.9} & \textbf{98.3} & \textbf{87.9}  \\
      \bottomrule
    \end{tabular}%
   }
   \end{resizebox}
   \linespread{0.9}
   \caption{Comparisons of different decoding algorithms on three test datasets. For a fair comparison, we ensure the false alarms are similar across different systems. Beam-search implementation of TDT is not available at the time of writing.}
  \label{table:different_decoding_algorithms}
% \vspace{-7pt}
\end{table}

% Average speed-up is measured on 100 utterances selected from each dataset, against greedy search baseline.

\begin{table*}[t]
  \centering
  \newcolumntype{S}{>{\small}c}
  \begin{resizebox}{1.8\columnwidth}{!}
  {
    \begin{tabular}{ S S S S S S S S S S S }
      \toprule
      \multirow{2}*{\textbf{Model}} & \multirow{2}*{$\bm{\mathcal{D}_\text{max}}$ } & \multicolumn{3}{c}{\textbf{Macro-recall}} & \multicolumn{3}{c}{\textbf{Rel. S. Speed-up}} & \multicolumn{3}{c}{\textbf{Rel. R. Speed-up}}\\
      \cmidrule(lr){3-5}  \cmidrule(lr){6-8}  \cmidrule(lr){9-11} 
      ~ & ~ & Snips & test-clean & test-other & Snips &  test-clean & test-other & Snips &  test-clean & test-other \\
       
      \midrule
       RNN-T KWS  & - & 98.1 & 99.0 & 90.2 & - & - & - & - & - & - \\ 
      \midrule

      \multirow{5}*{TDT-KWS} 
         & 2 & 98.2 & 97.9 & 89.3 & 1.44X & 1.57X & 1.57X & 1.39X & 1.53X & 1.54X \\
      ~  & 4 & 98.0 & 98.5 & \textbf{91.9} & 3.20X & 2.03X & 1.98X & 2.88X & 1.95X & 1.90X \\
      ~  & 6 & 97.7 & \textbf{99.1} & 90.3 & 3.63X & 2.06X & 2.03X & 3.20X & 1.97X & 1.95X \\
      ~  & 8 & 96.2 & 98.8 & 90.9 & 4.09X & 2.06X & 2.00X & 3.52X & 1.97X & 1.92X \\
      ~  & 10 & \textbf{98.6} & 98.7 & 89.5 & \textbf{4.19X} & \textbf{2.15X} & \textbf{2.05X} & \textbf{3.58X} & \textbf{2.04X} & \textbf{1.97X} \\
      \bottomrule
    \end{tabular}%
   }
   \end{resizebox}
   \linespread{0.9}
    % \caption{Performance comparison between RNN-T KWS and TDT-KWS under FAR = $0.02/h$ on ``Hey Snips" dataset \protect\footnotemark[1]  and under averaged FAR = $0.67/h$ on LibriKWS-20 dataset \protect\footnotemark[2]. Relative speed-up is measured against the RNN-T model. $\bm{\mathcal{D}_{max}}$ is maximum duration which can be skipped.}
    \caption{Performance comparison between RNN-T KWS and TDT-KWS under FAR = $0.02/h$, ($2/97h$) on ``Hey Snips" dataset and under averaged FAR = $0.67/h$, ($2/3h$) on LibriKWS-20 dataset. Relative speed-up is measured against the RNN-T model. $\bm{\mathcal{D}_\text{max}}$ is a hyper-parameter for TDT models representing the maximum duration that can be skipped.}
  \label{table:snips_libriKWS_results}
% \vspace{-7pt}
\end{table*}

\subsection{Evaluation Metrics}
To evaluate the performance of different models, we report the recall at a specific false alarm rate (FAR). For LibriKWS-20, which includes 20 keywords, it is typical to use \emph{macro-recall} and averaged FAR across all keywords. We measure inference speed in two metrics: 1. the overall inference speed-up of the system and 2. the execution speed-up of the inference module. We denote them as \emph{relative running speed-up} and \emph{relative search speed-up}, respectively.

%In a KWS system, achieving a high wake-up rate (recall) while maintaining a sufficiently low false alarm rate is crucial. To evaluate the performance of such systems, we present the recall at a specific false alarm rate (FAR). For LibriKWS-20, which includes 20 keywords in each testset, it is typical to use macro-recall and averaged FAR across all keywords in each testset. Additionally, two perspectives for measuring speed are considered: the overall inference time of the system and the execution time of the inference module. The relative running speed-up refers to the improvement in the overall inference time of the system compared to a baseline, while the relative search speed-up specifically focuses on the execution time of the inference module. We denote them as relative running speed-up and relative search speed-up, respectively.

% We also demonstrate the performance of different systems by comparing the region of interest of on the receiver operating characteristic (ROC) curves.

% \begin{figure}[t]
%   \centering
%     \includegraphics[width=1.0\linewidth]{snips_different_snr.png}
%     \vspace{-1.5em}
%     \caption{\it todo}
%     \vspace{-1.5em}
%     \label{fig:rnnt_kws_decoding}
% \end{figure}

\section{results and analysis}
\label{sec:rec_and_ana}

\subsection{Decoding Algorithm Comparison: ASR VS KWS-specific}

We compare our proposed KWS-decoding algorithm with ASR decoding methods for KWS tasks in terms of \emph{macro recall} and speed. As the results show in \Cref{table:different_decoding_algorithms}, our method significantly outperforms ASR decoding algorithms, both in greedy or beam-search modes. This improvement in performance can be attributed to the fact that our decoding algorithm restricts the search space to only the keyword, rather than all possible decoding sequences, which makes our algorithm achieve better results. As a result, we only use KWS-decoding algorithms for all subsequent experiments.

%which makes our algorithm more efficient and achieve better results. As a result, we only use KWS-decoding algorithms for all subsequent experiments.

\subsection{Model Performance: TDT VS RNN-T}

\begin{figure}[t]
  \centering
    \includegraphics[width=\linewidth]{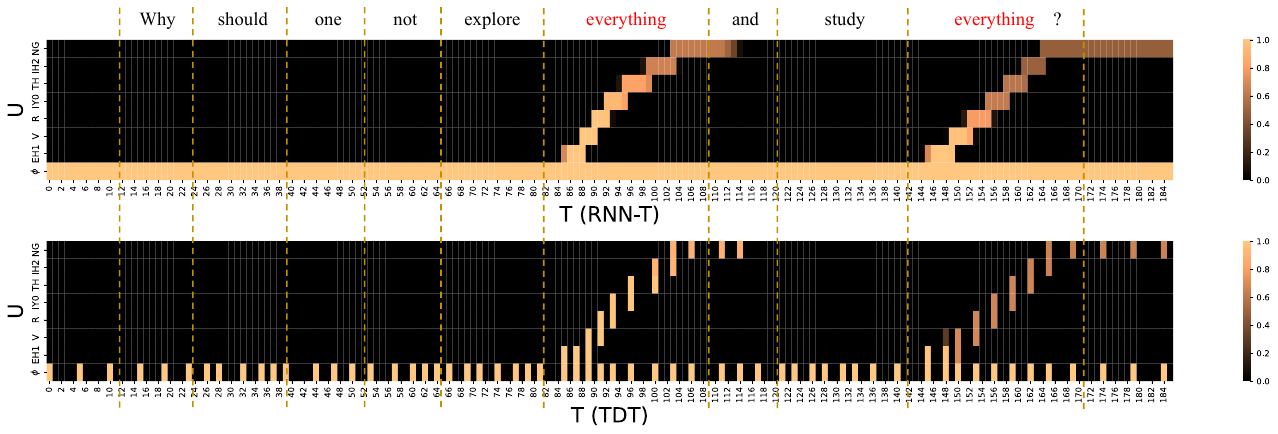}
    \vspace{-1.5em}
    \caption{Heatmaps of the wake-up score at each (t,u). The utterance is picked from the test-clean dataset, and the keyword is {\it everything}. The vertical yellow dashed lines represent the boundary information derived from force-alignments. Please zoom in to view the details.}
    \vspace{-1.5em}
    \label{fig:rnnt_kws_decoding}
\end{figure}

In \Cref{table:snips_libriKWS_results}, the performance of RNN-T KWS and TDT-KWS systems with different maximum duration skipping value, $\mathcal{D}_\text{max}$, is presented. The results show that TDT-KWS can achieve comparable or better performance than RNN-T KWS. More importantly, all TDT-KWS models achieve significant speed-up compared to the RNN-T KWS system. This can also be shown by the examples in \Cref{fig:rnnt_kws_decoding}, where TDT obviously has less search frequency. Moreover, the speed-up becomes even more pronounced as the value of $\mathcal{D}_\text{max}$ increases. The model still achieves superb results for Hey Snips when $\mathcal{D}_\text{max}$ is large. But for LibriKWS-20,  the best performance is observed when $\mathcal{D}_\text{max}$ is set to 4 or 6, and it gradually degrades as $\mathcal{D}_\text{max}$ increases beyond these values. These findings highlight the trade-off between performance and computational efficiency when using TDT-KWS systems with different $\mathcal{D}_\text{max}$ values.

The observed results align with intuitive expectations. In the case of the Hey Snips dataset, each positive test utterance consists of silence segments and the single phrase ``Hey Snips", whose pattern is relatively straightforward for the model to learn. Consequently, even with a significant number of skipped frames, the model can confidently identify frames containing crucial phonetic information. Therefore, the model performs well regardless of a larger value for $\mathcal{D}_\text{max}$. On the other hand, for the more challenging LibriKWS-20 dataset, when $\mathcal{D}_\text{max}$ is not very large, the model improves its performance by selectively disregarding interfering frames and focusing on frames that contain essential phonetic information, contributing to the enhancement of the model's performance. However, as $\mathcal{D}_\text{max}$ increases significantly, the model may skip specific tokens that should have been predicted, leading to a performance degradation. 

% By selectively skipping frames, the TDT model achieves a substantial speed improvement compared to the RNN-T. It achieves approximately 3 to 4 times speed-up on the Hey Snips test set and about 2 times on the two test sets within LibriKWS-20. 

%These findings indicate that the TDT model balances performance and speed, achieving notable speed improvements while maintaining competitive performance on both datasets.

% \vspace{-0.3cm} 

\begin{figure}[t]
  \centering
    \includegraphics[width=0.8\linewidth]{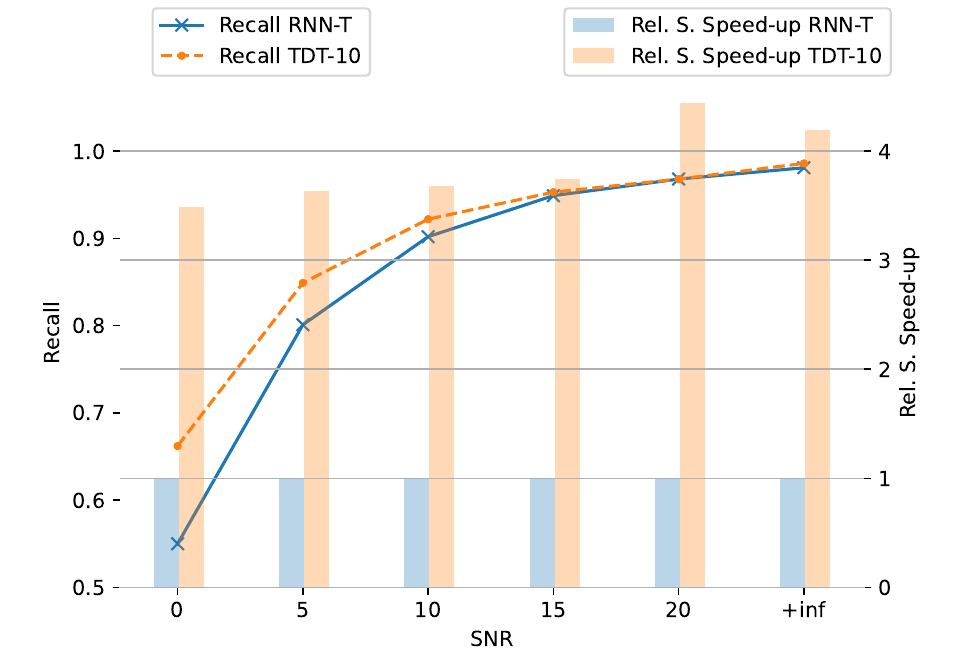}
    \vspace{-1.5em}
    \caption{Recall and inference speed comparison between RNN-T KWS and TDT-KWS at different SNR. SNR=+inf means no noise is added.}
    \vspace{-1.5em}
    \label{fig:results_in_noise}
\end{figure}

% \footnotetext[1]{The false alarm dataset for ``Hey Snips" is about \textbf{97} hours. We choose the number of false alarm is \textbf{2} ($FAR\approx0.02$ per hour) to present.}
% \footnotetext[2]{The false alarm datasets are both about \textbf{3} hours for LibriKWS-20. We choose the number of false alarm is \textbf{2} ($FAR\approx0.67$ per hour) to present.}

\subsection{Noise Robustness}
% In this section, we run decoding on Hey Snips dataset augmented with WHAM! noise in different SNRs to explore the robustness of RNN-T and TDT-KWS systems. The positive test samples are augmented with noise samples in 0, 5, 10, 15, and 20 SNRs, and we do not retrain models with augmented data. We report the recall and the relative searching speed-up for RNN-T KWS and TDT-KWS with $\mathcal{D}_\text{max}$ = 10 in \Cref{fig:results_in_noise}. As the noise get louder, the performance gap between the RNN-T KWS and the TDT-KWS widens further, demonstrating the TDT model's stronger robustness to noise. Additionally, TDT-KWS exhibits consistent speed improvements relative to the RNN-T KWS across different SNRs.

In this section, the RNN-T KWS and TDT-KWS systems are evaluated for their robustness to noise by running decoding on the Hey Snips dataset augmented with WHAM! noise at different SNRs. The positive test samples are augmented with noise samples in 0, 5, 10, 15, and 20 SNRs, without retraining the models on the augmented data. The results, as shown in \Cref{fig:results_in_noise}, indicate that as the noise level increases, the performance gap between the RNN-T KWS and the TDT-KWS widens further. Moreover, the TDT-KWS system consistently exhibits speed improvements relative to the RNN-T KWS system across different SNRs. This indicates that the TDT model not only achieves better performance in noise but also maintains its efficiency by providing consistent speed enhancements. Overall, these findings emphasize the TDT-KWS system's ability to effectively handle noise and exhibit rapid search speed, making it a promising choice for KWS in noisy environments.

% Our proposed method has two advantages: (1) Superior performance. The method we proposed achieved better performance on all three test datasets. (2) Our method is more flexible, allowing for the adjustment of thresholds to find the optimal balance between recall and false alarm rate. But the ASR decoding algorithm like greedy-search cannot adjust this dynamically.

% \begin{figure}[h]
%   \centering
%     \includegraphics[width=\linewidth]{wake_score_heatmap.pdf}
%     \vspace{-1.5em}
%     \caption{Heatmaps of wake-up score at each (t,u). The utterance is picked from test-clean dataset, and the keyword is {\it Everything}. The vertical yellow dashed lines represent the boundary information derived from force-alignments}
%     \vspace{-1.5em}
%     \label{fig:heatmap}
% \end{figure}

% We present the score heatmaps for both RNN-T and TDT-KWS with $\mathcal{D}_\text{max}$ = 6 during the inference of the same sentence in \Cref{fig:heatmap}. From the figure, we can observe several patterns: (1) Both models are capable of obtaining complete score paths at the positions where the keywords occur, thereby activating the wake-up state. (2) The TDT model achieves higher scores at the two wake-up points, which helps in differentiating between positive and negative examples and obtaining better classification boundaries. (3) The vertical yellow dashed lines represent the boundary information for each word. We can see that the models' wake-up positions are close to the end of the wake-up words, indicating a relatively low system latency.

\section{conclusions}
In this paper, we propose TDT-KWS, which applies Token-and-Duration Transducers (TDT) to KWS tasks. Our experiments show TDT-KWS not only outperforms the RNN-T KWS in terms of performance but also exhibits a significant improvement in inference speed. Additionally, TDT-KWS showcases enhanced robustness in noisy environments. We also propose an efficient KWS-specific decoding algorithm for Transducers in continuous streaming scenarios. It is more suitable for the KWS task, demonstrating superior performance to conventional ASR decoding methods. In the future, we will further explore the potential of TDT-KWS in more complex acoustic environments, and continue to optimize our decoding algorithm.

% The proposed search algorithm is more suitable for KWS task compared to ASR decoding algorithms, with the following strengths: 1. The model achieves superior performance than the existing ASR search methods. 2. The search algorithm is efficient, and the running speed is much faster than ASR decoding, which is crucial for achieving low latency. In the future, we will further explore the potential of TDT-KWS in more complex acoustic environments, and continue to optimize our decoding algorithm.

%The searching speed is much more faster than ASR decoding, which is critical for KWS to achieve low latency.
% Body

\newpage
% \linespread{0.9}
\bibliographystyle{IEEEbib}
\bibliography{main}

\end{document}